\documentclass{revtex4}
\usepackage[dvips]{graphicx}

\begin{document}
\title{To the origin of large reduction of the effective moment\\
in Na$_{2}$V$_{3}$O$_{7}$}
\author{Z. Ropka}
\affiliation{Center for Solid State Physics, S$^{nt}$Filip 5,
31-150 Krakow, Poland}
\author{R.J. Radwanski}
\homepage{http://www.css-physics.edu.pl}
\email{sfradwan@cyf-kr.edu.pl}
\affiliation{Center for
Solid State Physics, S$^{nt}$Filip 5, 31-150 Krakow, Poland,\\
Institute of Physics, Pedagogical University, 30-084 Krakow,
Poland}

\begin{abstract}
We have shown that the observed large reduction of the effective moment in Na%
$_{2}$V$_{3}$O$_{7}$ may be caused by conventional crystal-field
interactions and the intra-atomic spin-orbit coupling of the V$^{4+}$ ion.
The fine discrete electronic structure of the 3$d^{1}$ configuration with
the weakly-magnetic Kramers-doublet ground state, caused by the large
orbital moment, is the reason for anomalous properties of Na$_{2}$V$_{3}$O$%
_{7}$. Moreover, according to the Quantum Atomistic Solid-State Theory
(QUASST) Na$_{2}$V$_{3}$O$_{7}$ is expected to exhibit pronounced
heavy-fermion phenomena at low temperatures.


\end{abstract}
\maketitle
Gavilano {\it et al.} \cite{1} have discovered that the
effective moment of the V$^{4+}$ ion in Na$_{2}$V$_{3}$O$_{7}$ is
reduced by 1 order of magnitude upon reducing the temperature from
100 to 10 K. The moment reduction is inferred from the
experimentally measured temperature dependence of the magnetic
susceptibility. After taking into account the diamagnetic
contribution the inverse susceptibility shows in the temperature
range 100-300 K a straight line behaviour with the effective
moment p$_{eff}$
of 1.9 $\mu _{B}$ per V ion. A straight line between 20 and 1.9 K implies $%
p_{eff}$ of 1 order of magnitude smaller. Gavilano {\it et al.} provide an
explanation that ''the reduction of the effective magnetic moment is most
likely due to a gradual process of moment compensation via the formation of
singlet spin configurations with most but not all of the ions taking part in
this process. This may be the result of antiferromagnetic interactions and
geometrical frustration.'' They further conjectured ''the compensation of
eight out of the nine V spins ...'' in order to reproduce the observed
reduction of the effective moment by 1 order of magnitude. It is worth to
add that Na$_{2}$V$_{3}$O$_{7}$ shows no sign of the magnetic order down to
1.9 K.

The aim of this Letter is to propose an explanation for this reduction of
the effective moment in Na$_{2}$V$_{3}$O$_{7}$. We can reproduce very well
the observed temperature dependence of the paramagnetic susceptibility by
considering the electronic structure associated with the V$^{4+}$ ion (3$%
d^{1}$ configuration) under the action of the crystal field (CEF)\ taking
into account the spin-orbit (s-o) coupling. It turns out that, despite of
the Kramers doublet ground state, a state with quite small magnetic moment
can be obtained as an effect of the spin-orbit coupling, that involves quite
large orbital moment. In our present explanation we have been oriented by
our earlier calculations for the V$^{4+}$ ion presented in Refs \cite{2,3}.

We assume that one 3$d$ electron in the V$^{4+}$ ion is described by quantum
numbers $L$=2 and $S$=1/2. The ground term $^{2}D$ is 10-fold degenerated.
Its degeneracy is removed by the intra-atomic spin-orbit interactions and in
a solid by crystal-field interactions. This situation can be exactly traced
by the consideration of a single-ion-like Hamiltonian

\begin{center}
$H_{d}=H_{CF}^{octa}+H_{s-o}+H_{CF}^{tr}+H_{Z}=B_{4}(O_{4}^{0}+5O_{4}^{4})+%
\lambda L\cdot S+B_{2}^{0}O_{2}^{0}+\mu _{B}(L+g_{e}S)\cdot {\bf B}\,\;\;\;$
\end{center}

in the 10-fold degenerated spin+orbital space. The last term allows to
calculate the influence of the external magnetic field {\bf B} and enables
calculations of the paramagnetic susceptibility. Such type of the single-ion
Hamiltonian has been widely used in analysis of electron-paramagnetic
resonance (EPR)\ spectra of 3$d$-ion doped systems \cite{4,5}. Here we use
this Hamiltonian for systems, where the 3$d$ ion is the full part of the
crystal. Although we know that the local symmetry in Na$_{2}$V$_{3}$O$_{7}$
is quite complex we approximate, for simplicity, the CEF\ interactions by
considering dominant octahedral interactions with a trigonal distortion. For
the octahedral crystal field we take $B_{4}$= +200 K. The sign ''+'' comes
up from {\it ab initio} calculations for the ligand octahedron. The
spin-orbit coupling parameter $\lambda _{s-o}$ is taken as +360 K, as in the
free V$^{4+}$ ion \cite{4}.

The resulting electronic structure of the 3$d^{1}$ ion contains 5 Kramers
doublets separated in case of the dominant octahedral CEF interactions into
3 lower doublets, the $T_{2g}$ cubic subterm, and 2 doublets, the $E_{g}$
subterm, about 2 eV above (Fig. 1). The $T_{2g}$ subterm in the presence of
the spin-orbit coupling is split into lower quartet and excited doublet
(Fig. 1.2). Positive values of the trigonal distortion parameter $B_{2}^{0}$
yields the ground state that has small magnetic moment (Fig. 1.3). For $%
B_{2}^{0}$ = +6 K the ground state moment amounts to $\pm 0.15$ $\mu _{B}$.
It is composed from the spin moment of $\pm 0.42$ $\mu _{B}$ and the\
orbital moment of $\mp 0.27$ $\mu _{B}$ (antiparallel). The sign $\pm $
corresponds to 2 Kramers conjugate states. The excited Kramers doublet lies
at 38 K (3.3 meV) and is almost non-magnetic - its moment amounts to $\pm
0.03$ $\mu _{B}$ only (=$\pm 1.03$ $\mu _{B}+2\cdot (\mp 0.50$ $\mu _{B})$)
due to cancellation of the spin moment by the orbital moment. So small and
different moments for subsequent energy levels is an effect of the
spin-orbit coupling and distortions.
\begin{figure}[ht]
\includegraphics[width = 8 cm]{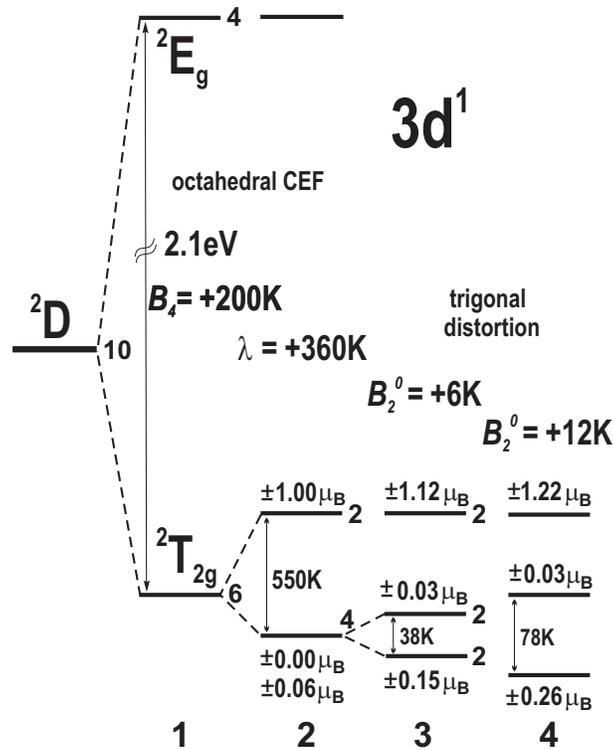}
\caption{The localized states of the 3$d^{1}$ configuration in the
V$^{4+}$ ion under the action of the crystal field and spin-orbit
interactions originated from the 10-fold degenerated $^{2}D$ term;
(1) the splitting of the $^{2}D$ term by the octahedral CEF\ surroundings with $B_{4}$=+200 K, $%
\protect\lambda _{s-o}$ =0; (2) the splitting by the combined
octahedral CEF and spin-orbit interactions; (3) and (4) the effect
of the trigonal distortions. The states are labelled by the
degeneracy in the spin-orbital space and the value of the magnetic
moment.}
\end{figure}

Having the electronic structure, the state energies and eigenfunctions, we
calculate the Helmholtz free energy $F(T,B)$. By definition, known from
statistical physics, we calculate the paramagnetic susceptibility, as $\chi
(T)$ $=$ $-\partial ^{2}F(T,B)/\partial B^{2}$, and its temperature
dependence. In Figs 2 and 3 the calculated results for the paramagnetic
susceptibility are shown for different physical situations of the V$^{4+}$
ion: line (1) for the purely octahedral crystal field; line (2) shows $\chi
(T)$ in extra presence of the spin-orbit coupling; lines (3), (4), (5) in
extra presence of trigonal distortions $B_{2}^{0}$ of different strength.
One can see that all these curves are completely different from the $S$=1/2
behaviour, expected for a free spin. The V$^{4+}$ ion with one 3$d$ electron
is usually treated as $S$=1/2 system i.e. with the spin-only magnetism and
with taking into account the spin degree of freedom only. The neglect in the
current literature of the orbital moment is consistent with the
widely-spread conviction that the orbital magnetism plays rather negligible
role due to the quenching of the orbital moment for 3$d$ ions. Na$_{2}$V$%
_{3} $O$_{7}$ is an example of numerous compounds in which the $S$=1/2
behavior in the temperature dependence of the paramagnetic susceptibility is
drastically violated (CaV$_{4}$O$_{9}$, MgVO$_{3}$, (VO)$_{2}$P$_{2}$O$_{7}$%
, ...). It is seen in the substantial departure of $\chi (T)$ from the Curie
law at low temperatures. The spin-orbit coupling and small distortion of the
local surroundings of the V$^{4+}$ ion causes drastic change of the slope of
the $\chi ^{-1}(T)$ plot below 100 K (Fig. 3, lines (2) and (5)). Such
change of the slope is usually treated as the change of the effective
moment. In Fig. 4 the temperature dependence of the effective moment per the
V ion, calculated as $p_{eff}=\sqrt{3\chi T}$, are presented for different
physical situations shown in Fig. 2. The strong temperature dependence is
seen below 100 K, Fig. 4 lines (2) and (5). It means, that the strong
temperature dependence results from the spin-orbit coupling and distortions.
It is worth to add that negative values of the $B_{2}^{0}$ yields the
temperature dependence of the susceptibility with a broad maximum due to the
almost non-magnetic ground state and excited magnetic doublet, similarly as
we calculated in Ref. \cite{2}.
\begin{figure}[ht]
\includegraphics[width = 10 cm]{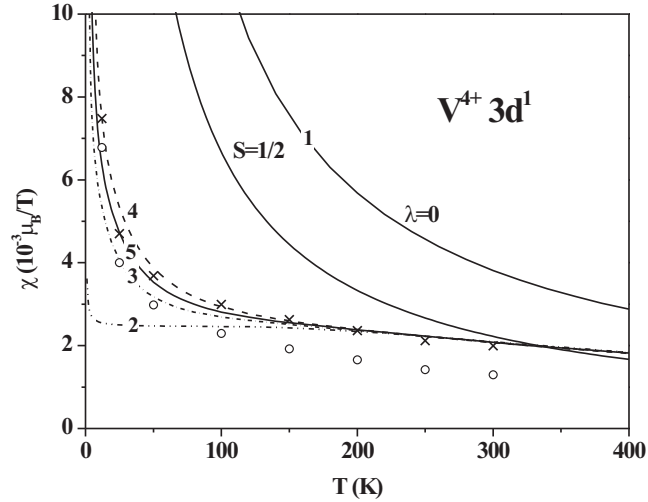}
\caption{The calculated temperature dependence of the atomic-scale
paramagnetic susceptibility $\protect\chi (T)$ for the 3$d^{1}$
configuration in the V$^{4+}$ ion for different physical
situations: \ line
(1) - $\protect\chi (T)$ for the purely octahedral crystal field with $B_{4}$%
=+200K ($\protect\lambda _{s-o}$= 0); line (2) - in combination
with the spin-orbit coupling $\protect\lambda _{s-o}$= +360 K;
lines (3) and (4) show the influence of the off-cubic trigonal
distortions $B_{2}^{0}$=+6 K (3) and $B_{2}^{0}$=+12 K (4); curve
(5), with $B_{2}^{0}$=+9 K, reproduces very well measured
experimental data (o, after Refs \protect\cite{1,6}) after taking
into account a diamagnetic term (x).}
\end{figure}
\begin{figure}[ht]
\includegraphics[width = 10 cm]{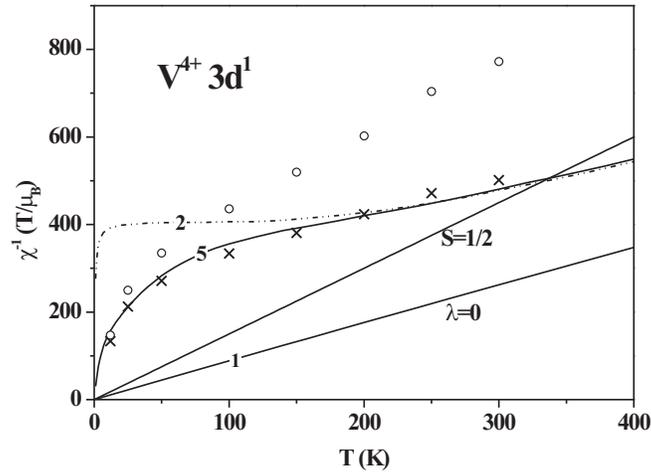}
\caption{The calculated temperature dependence of the atomic-scale
paramagnetic susceptibility shown in the $\protect\chi ^{-1}$ {\it
vs} $T$ plot for the 3$d^{1}$ configuration in the V$^{4+}$ ion
for different physical situations defined in Fig. 1 and in the
main text. The significant departure from the Curie law as well as
from the S=1/2 behavior is seen due to the spin-orbit coupling and
distortions.}
\end{figure}
\begin{figure}[ht]
\includegraphics[width = 10 cm]{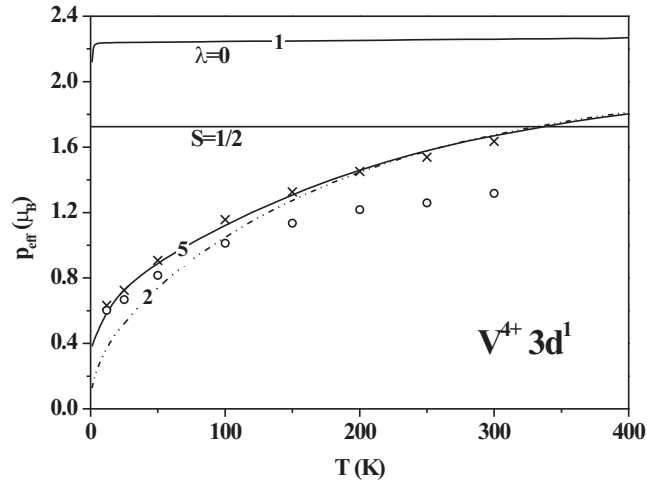}
\caption{Temperature dependence of the effective magnetic moment
$p_{eff}$ calculated from the Curie law for the 3$d^{1}$
configuration in the V$^{4+}$ ion for different physical
situations defined in Fig. 1 and in the main text. The significant
reduction from a value of 1.73 $\protect\mu _{B}$, expected for
the free spin S=1/2, is seen due to the spin-orbit coupling and
distortions.}
\end{figure}

For comparison we show in Figs 2 and 3 the experimental data
(empty circles) taken from Refs \cite{1,6} for the measured
paramagnetic susceptibility. These data have been used for
calculations of the effective moment from the Curie law - they are
shown in Fig. 4 as empty circles. The data, shown as
''x'' in Figs 2-4 have been obtained from measured experimental data of Refs %
\cite{1,6} by taking into account a diamagnetic term $\chi _{o}$ of -0.0007 $%
\mu _{B}$/T V-ion (= -0.0004 emu/mol V). These data (x) coincide well with
our calculations shown as line 5 obtained for a set of parameters: $B_{4}$=
+200 K, $\lambda _{s-o}$ = +360 K and $B_{2}^{0}$ = +9 K. For this set of
parameters the energy level scheme contains 2 excited doublets at 58 \ and
580 K and the ground-state moment amounts to 0.21 $\mu _{B}$. We treat this
coincidence as not fully physically relevant owing to the much more complex
local symmetry of the V$^{4+}$ ion in Na$_{2}$V$_{3}$O$_{7}$, a large
uncertainty in the evaluation of the diamagnetic term and of the
paramagnetic susceptibility measured on a polycrystalline sample. But we
take the reached agreement as strong indication for the existence of the
fine electronic structure in Na$_{2}$V$_{3}$O$_{7}$, originating from the V$%
^{4+}$ ion, determined by crystal-field and spin-orbit interactions.
Extremely important is that we can reproduce not only the overall $\chi (T)$
dependence in the full measured temperature range but we also reproduce its
absolute value.

In conclusion, we argue that the experimentally-observed temperature
dependence of the paramagnetic susceptibility of Na$_{2}$V$_{3}$O$_{7}$ can
be explained in single-ion approach, extended to the Quantum Atomistic
Solid-State Theory QUASST \cite{7,8}, taking into account the crystal-field
interactions, the intra-atomic spin-orbit coupling and the orbital
magnetism. It is further indication for the high physical adequacy of the
QUASST\ conjecture that the electronic and magnetic properties of the 3{\it %
d-}ion containing compounds are determined by the fine electronic structure
of the 3{\it d} ion. Moreover, we would like to point out that according to
QUASST the Kramers spin-like degeneracy of the ground state, not removed
down to 1.9 K, has to be removed somewhere - we expect it to occur at very
low temperatures. Thus our theory predicts Na$_{2}$V$_{3}$O$_{7}$ to exhibit
heavy-fermion-like properties in the specific heat at ultra-low temperatures.


\begin{thebibliography}{}
\bibitem{1} J. L. Gavilano, D. Rau, S. Mushkolaj, H. R. Ott, P. Millet, and
F. Mila, Phys. Rev. Lett. {\bf 90,} 167202 (2003).
\bibitem{2} R. J. Radwanski, R. Michalski, and Z. Ropka, Physica B {\bf %
312-313} 628 (2002).
\bibitem{3} Z. Ropka, R. Michalski, and R. J. Radwanski, {\it Anomalous
temperature dependences of the susceptibility for the one-3d-electron cation}%
, http://xxx.lanl.gov/abs/cond-mat/9907141.
\bibitem{4} A. Abragam and B. Bleaney, {\it Electron Paramagnetic Resonance
of Transition Ions} (Clarendon, Oxford, 1970).
\bibitem{5} C. Ballhausen: {\it Ligand Field theory }(McGraw-Hill, 1962); W.
Low: {\it Paramagnetic Resonance in Solids} (Academic, 1960).
\bibitem{6} J. L. Gavilano, D. Rau, S. Mushkolaj, H. R. Ott, P. Millet, and
F. Mila, Physica B {\bf 312-313} 622 (2002).
\bibitem{7} R. J. Radwanski, R. Michalski, and Z. Ropka, Acta Phys. Polonica
B {\bf 31} 3079 (2000).
\bibitem{8} R. J. Radwanski and Z. Ropka, {\it Quantum Atomistic Solid State
Theory}, http://xxx.lanl.gov/abs/cond-mat/0010081.
\end{thebibliography}
\end{document}